\documentclass[11pt,a4paper]{article}
\usepackage{jheppub}
\usepackage{slashed}
\usepackage{amsmath, amssymb, graphicx, physics, dsfont,array}
\usepackage{orcidlink}
\usepackage[normalem]{ulem}
\usepackage{comment}
\usepackage{multirow}
\usepackage{bm}
\usepackage{graphicx}

\newcommand{\as}{\alpha_s}
 
\newcommand{\be}{\begin{equation}}
\newcommand{\ee}{\end{equation}}
\newcommand{\bes}{\begin{subequations} \begin{align} }
\newcommand{\ees}{\end{subequations}\end{align} }
\newcommand{\bea}{\begin{eqnarray}}
\newcommand{\eea}{\end{eqnarray}}

\newcommand{\eq}[1]{Eq.~\eqref{eq:#1}}

\renewcommand{\sec}[1]{Sec.~\ref{sec:#1}}

\newcommand{\fig}[1]{Fig.~\ref{fig:#1}}

\newcommand{\tab}[1]{Table~\ref{tab:#1}}

\def\bal#1\eal{\begin{align}#1\end{align}}

\newcounter{RSQ}

\newcounter{MSQ}

\title{\boldmath $J/\psi$ Production Within Jets at the EIC}

\author[a]{Yunlu Wang\orcidlink{0009-0003-1313-1093}}
\author[b]{Hee Sok Chung\orcidlink{0000-0003-1628-9315}}
\author[a]{Taewook Ha\orcidlink{000-0002-5725-4007}}
\author[a,c]{Daekyoung Kang\orcidlink{0000-0002-9145-6913}}
\author[d]{U-Rae Kim\orcidlink{0000-0003-4473-4834}}

\affiliation[a]{Key Laboratory of Nuclear Physics and Ion-beam Application (MOE) and Institute of Modern Physics, Fudan University, Shanghai, China 200433}
\affiliation[b]{Department of Mathematics and Physics, Gangneung-Wonju National University, Gangneung 25457, Korea}
\affiliation[c]{Department of Physics, Korea University, Seoul 02841, Korea}
\affiliation[d]{Department of Physics, Korea Military Academy, Seoul 01805, Korea}

\emailAdd{yunluwang20@fudan.edu.cn}
\emailAdd{taewookha@fudan.edu.cn}
\emailAdd{dkang@fudan.edu.cn}
\emailAdd{kim87@kma.ac.kr}
\emailAdd{heesokchung@gwnu.ac.kr}

\abstract{
We present theoretical predictions for the transverse-momentum distribution of $J/\psi$ produced within jets at the upcoming Electron-Ion Collider (EIC). 
Utilizing the semi-inclusive fragmenting jet function (FJF) framework, our calculation achieves next-to-leading order (NLO) accuracy in the strong coupling and leading-logarithmic (LL) accuracy by resumming both collinear and threshold logarithms.
In contrast to the gluon-dominated regime of the LHC, EIC photoproduction is characterized by an enhanced quark-initiated component, offering a complementary probe into the charmonium production mechanism governed by nonperturbative long-distance matrix elements (LDMEs). We examine the impact of representative LDME sets, demonstrating the EIC's distinct discriminating power for the mechanisms.
We find that quark contributions are particularly significant in the small momentum fraction region. This region is also shown to be sensitive to both the jet radius $R$ and the experimental muon identification criteria for the $J/\psi \to \mu^+\mu^-$ decay channel.
These findings establish quarkonium-in-jet observables at the EIC as a vital, independent probe for constraining the production mechanisms and advancing our understanding of heavy quarkonium formation. 
}

\begin{document}
\maketitle
\section{Introduction} 

Since its discovery over 50 years ago~\cite{Augustin:1974xw,Aubert:1974js}, $J/\psi$ production has been extensively studied to understand its production mechanism and to serve as a tool for probing QCD interactions, both theoretically and experimentally.
Despite the significant success of nonrelativistic quantum chromodynamics (NRQCD) \cite{Bodwin:1994jh}—an effective theory for heavy quarkonium—in describing transverse momentum ($p_T$) distributions, several puzzles remain unresolved \cite{Braaten:2002fi, Braaten:1999qk}. In particular, inclusive $p_T$ distributions do not provide sufficient degrees of freedom to accurately constrain the long-distance matrix elements (LDMEs) that govern the nonperturbative transition of a heavy quark-antiquark pair into a physical quarkonium state. Consequently, the LDME sets reported in the literature \cite{Butenschoen:2011yh,Gong:2012ug,Han:2014jya,Zhang:2014ybe,Shao:2014yta,Bodwin:2015iua,Feng:2018ukp,Brambilla:2021abf,Butenschoen:2022qka,Chung:2024jfk} either suffer from large uncertainties or exhibit significant mutual tensions. To address these issues, a more differential observable, such as the transverse-momentum distribution of quarkonia within jets, has been proposed as a way to discriminate between competing production mechanisms \cite{Baumgart:2014upa, Bain:2016rrv, Kang:2017yde}.

Transverse momentum distributions of charmonia within jets have been measured at hadron colliders~\cite{LHCb:2017llq, CMS:2021puf, Yang:2021smr, LHCb:2024ybz}. 
Theoretical studies utilizing the fragmenting jet function (FJF) framework \cite{Bain:2017wvk, Copeland:2025osx} have successfully described LHCb charmonium measurements and demonstrated a significant potential to discriminate between competing LDME sets. 
Recently, more refined theoretical predictions \cite{Wang:2025drz} have been developed, incorporating threshold resummation~\cite{Chung:2024jfk} and a factorization formalism that is better aligned with experimental kinematic conditions~\cite{Kang:2016ehg}.
However, a definitive determination of the LDMEs still requires further refinements in theoretical calculations and more extensive analysis.

In this regard, additional constraints and complementary information are essential for a definitive extraction of these matrix elements. Recent studies \cite{Wang:2025drz, LHCb:2024ybz} have shown that examining distributions in the dual space of charmonium $p_T$ and jet $p_T$ provides more comprehensive information regarding the production mechanism. Independent experiments at electron-proton facilities, such as the upcoming Electron-Ion Collider (EIC) \cite{AbdulKhalek:2021gbh, Boer:2024ylx}, provide an ideal alternative platform for measuring quarkonium production within jets.

Charmonium production in electron-proton ($ep$) and proton-proton ($pp$) collisions shares 
a similar factorization formula, utilizing the same FJFs but distinct partonic production cross sections (PPCSs). 
In $pp$ collisions, the gluon fusion process dominates and enhances gluon fragmentation, while quark fragmentation remains relatively suppressed. 
In contrast, the relative importance of quark fragmentation is significantly enhanced in $ep$ collisions.
This shift leads to different production mechanisms and the relative contributions of the long-distance matrix elements (LDMEs) differ from those observed in $pp$ collisions.
Therefore, measurements at the EIC provide information complementary to those performed at the LHC.

In this paper, we predict the distribution of the transverse momentum fraction, $z_{J/\psi}$, carried by the $J/\psi$ produced within a jet at the EIC. We employ the semi-inclusive Fragmenting Jet Function (FJF) formalism developed in Refs.~\cite{Kaufmann:2015hma,Kang:2016ehg}, which is formulated in the large-$p_T$ limit.
To enhance theoretical precision, we resum collinear logarithms associated with the hierarchies between the hard scale $p_T$, the jet scale $p_T R$, and the quarkonium mass scale $m_{J/\psi} \approx 2 m_c$. This resummation is achieved through the Dokshitzer–Gribov–Lipatov–Altarelli–Parisi (DGLAP) evolution of both the fragmentation functions (FFs) and FJFs~\cite{Gribov:1972ri, Lipatov:1974qm, Dokshitzer:1977sg, Altarelli:1977zs}. Furthermore, we implement threshold resummation to ensure the distribution is convergent and well-behaved near the kinematic threshold $z_{J/\psi} \to 1$. We select three representative sets of LDMEs  as inputs for our predictions to test the discriminating power of available LDME sets.
The resulting distributions are presented for various transverse-momentum intervals and jet radii. To provide a point of contrast, selected predictions are compared with corresponding results for LHCb.

The rest of this paper is organized as follows. In \sec{2}, we review the factorization formula, the fragmenting jet functions, and their resummation. In \sec{3-1}, we describe our numerical setup, including the PPCSs, the three LDME sets, and the feeddown contributions. In \sec{3-2}, we present our numerical results for the transverse-momentum fraction distributions. Finally, we summarize our findings in \sec{4}.

\section{Theoretical Framework}
\label{sec:2}

The hadron-in-jet process was originally explored and its factorization formulated for exclusive jet production \cite{Procura:2009vm, Jain:2011xz, Procura:2011aq}; this framework was later applied to quarkonium production~\cite{Baumgart:2014upa, Bain:2017wvk}. Subsequently, factorization for semi-inclusive jet processes~\cite{Kaufmann:2015hma, Kang:2016mcy, Kang:2016ehg}, which more closely align with experimental configurations, was introduced. In the following subsections, we review the relevant factorization for semi-inclusive FJFs and the associated resummation.

\subsection{Factorization formula}
\label{sec:2-2}

The factorization formulae for FJFs were developed in Refs.~\cite{Kaufmann:2015hma, Kang:2016mcy} for $pp$ collisions and in Ref.~\cite{Aschenauer:2019uex} for $ep$ collisions. In the limit of large jet transverse momentum, the factorized cross section for $J/\psi$ production inside a jet at the EIC is expressed as
\begin{equation}\label{eq:cs1}
\frac{d\sigma^{ep\to (\text{jet }J/\psi)X}}{dp_T\, d\eta\, dz_{J/\psi}} 
= \sum_i \int_{z_0}^1 dz \; 
\frac{d\hat{\sigma}_{ep\to iX}(p_T/z, \eta,\mu)}{dp_Td\eta} \; 
\mathcal{G}_i^{J/\psi}(z, z_{J/\psi}, p_TR, \mu),
\end{equation}
where the PPCSs $d\hat{\sigma}_{ep\rightarrow iX}/(dp_Td\eta)$ correspond to the production of an outgoing parton $i$, with the index $i$ running over all available partonic channels. The FJFs $\mathcal{G}_i^{J/\psi}$ describe the distribution of the $J/\psi$ inside a jet initiated by parton $i$. The parameters $p_T$, $\eta$, and $R$ represent the transverse momentum, rapidity, and radius of the jet, respectively, while $\mu$ is the factorization scale. The momentum-fraction variables $z$ and $z_{J/\psi}$ are defined as
\begin{align}
z=\frac{p_T}{\hat{p}_T}, \quad z_{J/\psi}=\frac{p^{J/\psi}_T}{p_T},
\end{align}
where $\hat{p}_T$ and $p_T^{J/\psi}$ are the transverse momenta of the parton $i$ and the $J/\psi$, respectively. The lower bound of the integration variable $z$ is $z_0=2p_T\cosh\eta/\sqrt{s}$, where $s$ is the square of the CM energy of the $ep$ system.

The PPCSs are expressed as a convolution of the parton distribution functions (PDFs) of the incoming electron and proton with the perturbatively calculable hard functions~\cite{Aversa:1988vb,Jager:2002xm}.
\begin{align}
\frac{d\hat{\sigma}_{ep\to iX}(p_T/z, \eta,\mu)}{dp_Td\eta}
=& 
\frac{2p_T}{s} \sum_{a,b}
\frac{1}{z^2}
\int^{1-(1-V)/z}_{VW/z} \!\!\!\!
\frac{dv}{v(1-v)} 
\int^1_{VW/(vz)} 
\frac{dw}{w}
\nonumber\\
&\qquad\times 
f_{a/e}(x_a,\mu) f_{b/p}(x_b,\mu)
H^i_{ab} (x_a, x_b; p_{T}/z, \mu)\,,
\end{align}
where $f_{b/p}(x_b,\mu)$ denotes the probability density of finding a parton $b$ carrying a momentum fraction $x_b$ of the proton. Similarly, $f_{a/e}(x_a,\mu)$ describes a parton $a$ carrying the momentum fraction $x_a$ of the electron.
The hard function $H^i_{ab}$ represents the partonic cross section for the production of parton $i$ in the collision of partons $a$ and $b$, with all other final-state partons integrated out.
The integration limits are expressed in terms of the hadronic variables $V$ and $W$, defined by
\begin{align}
V = 1- \frac{p_T}{\sqrt{s}}e^{-\eta}
\,,\qquad
W = \frac{p_T^2}{sV(1-V)}\,.
\end{align}  
The momentum fractions $x_a$ and $x_b$ are expressed in terms of the integration variables $v$ and $w$ as
\begin{align}
x_a = \frac{VW}{zvw}\,,\quad x_b=\frac{1-V}{z(1-v)} \,.
\end{align}

The effective electron PDF $f_{a/e}(x_a,\mu)$ is expressed as a convolution of the photon flux emitted by the electron and the parton distributions within the photon:
\begin{equation}
f_{a/e}(x_a,\mu) = \int_{x_a}^{1}\frac{dy}{y}\,
f_{\gamma/e}(y)\,
f_{a/\gamma}\!\left(\frac{x_a}{y}, \mu\right),
\end{equation}
where the Weizs\"acker--Williams photon spectrum~\cite{deFlorian:1999ge,Frixione:1993yw} is given by
\begin{equation}
f_{\gamma/e}(y) =
\frac{\alpha}{2\pi}
\left[
\frac{1+(1-y)^2}{y}\ln\!\frac{Q^2_{\text{max}}(1-y)}{m_e^2 y^2}
- \frac{2(1-y)}{y}
\right].
\end{equation}
Here, $\alpha$ is the electromagnetic coupling constant, $m_e$ is the electron mass, and $y = E_\gamma / E_e$ is the photon energy fraction. Note that the photon virtuality, defined as the negative squared photon momentum transfer ($Q^2 = -q^2$), is integrated up to a cutoff value $Q^2_{\text{max}}$.

The parton-in-photon distribution $f_{a/\gamma}(x_\gamma, \mu)$ depends on the momentum fraction of parton $a$ relative to the photon. For the direct process, where the photon interacts directly with a parton from the proton, the parton-in-photon distribution  corresponds to $a=\gamma$ and is given by
\begin{equation}
f_{a/\gamma}(x_\gamma,\mu) = \delta_{a\gamma}\, \delta(1-x_\gamma).
\end{equation}
In contrast, the photon can also resolve into partons, which then interacts with another parton from the proton. This is known as the resolved process and gives rise to the parton-in-photon PDF, which is an additional nonperturbative input~\cite{Gluck:1991jc}. Several parton-in-photon PDF sets are available in the literature~\cite{Schuler:1995fk, Gordon:1996pm, Gluck:1999ub, Aurenche:2005da}.

The hard functions $H_{ab}^i$ for the resolved processes are equivalent to hadroproduction cross sections and were computed at NLO in Refs.~\cite{Aversa:1988vb, Jager:2002xm}. Similarly, the hard functions for the direct processes have been obtained in Refs.~\cite{Aurenche:1986ff, Gordon:1994wu, Abelof:2016pby}.

\subsection{Fragmenting jet functions}
\label{sec:2-4}

The FJFs are further factorized into $J/\psi$ fragmentation functions as follows
\begin{equation}
\mathcal{G}_i^{J/\psi}(z, z_{J/\psi},p_TR, \mu) 
= \sum_j \int_{z_{J/\psi}}^1 \frac{dz_{J/\psi}'}{z_{J/\psi}'} \; 
\mathcal{J}_{ij}(z, z_{J/\psi}', p_TR, \mu) \; 
D_j^{J/\psi}\Big(\frac{z_{J/\psi}}{z_{J/\psi}'}, \mu\Big),
\end{equation}
where $\mathcal{J}_{ij}$ is the perturbatively calculable coefficient describing the production of parton $j$ inside a jet initiated by parton $i$, and $D_j^{J/\psi}$ is the FF of parton $j$ into the $J/\psi$. Since the jet scale $p_T R$ is much larger than $\Lambda_{\text{QCD}}$, this relation is valid up to power corrections of $\mathcal{O}(\Lambda_{\text{QCD}}^2 / p_T^2 R^2)$. The $J/\psi$ FFs are subsequently factorized within the NRQCD framework as~\cite{Braaten:1993rw,Bodwin:1994jh}
\begin{equation}\label{eq:FF-fac}
D_j^{J/\psi}(z, \mu_0) = \sum_{\mathcal{N}} d_{j\to [Q\bar Q(\mathcal{N})]}(z, \mu_0) \; 
\langle \mathcal{O}^{J/\psi}_{[Q\bar Q(\mathcal{N})]} \rangle,
\end{equation}
where the short-distance coefficient $d_{j\to [Q\bar Q(\mathcal{N})]}$ describes the fragmentation of parton $j$ into a nonrelativistic $Q\bar{Q}(\mathcal{N})$ pair at the scale $\mu_0 = 2m_c$ (with $m_c$ denoting the charm quark mass), and can therefore be calculated perturbatively. The nonperturbative LDME $\langle \mathcal{O}^{J/\psi}_{[Q\bar Q(\mathcal{N})]} \rangle$, defined at the scale $\mu_{\Lambda} \sim \Lambda_{\text{QCD}}$, describes the transition of the $Q\bar{Q}(\mathcal{N})$ pair into the $J/\psi$. The summation runs over the four dominant channels, with $\mathcal{N}$ spanning $^3S_1^{[1]}, {}^1S_0^{[8]}, {}^3S_1^{[8]}$, and $^3P_J^{[8]}$, where the superscripts $[1]$ and $[8]$ denote color singlet and color octet states, respectively.

\subsection{Resummation}
\label{sec:resum}

The hierarchy $2m_c < p_T R < p_T$ induces theoretical uncertainties associated with large logarithms of the form $\ln R$ and $\ln(p_T R / 2m_c)$ in the cross sections. These uncertainties can be reduced by resumming the logarithms through DGLAP evolution. At leading order in $\alpha_s$, the evolution equation for the FFs~\cite{Bodwin:2015yma} reads
\begin{equation}\label{eq:DGLAP}
\frac{d}{d \ln \mu^2}\binom{D_S}{D_g}=\frac{\alpha_s\left(\mu\right)}{2 \pi}\left(
\begin{array}{cc}
P_{q q} & 2 n_f P_{g q} \\
P_{q g} & P_{g g}
\end{array}
\right) \otimes\binom{D_S}{D_g}\,,
\end{equation}
where $D_S = \sum_{f=1}^{n_f} (D_{q_f} + D_{\bar{q}_f})$ is the singlet FF summed over quark flavor $f$, $P_{ij}$ are the splitting functions, and $n_f=5$ is the number of active flavors. 
The FF is evolved from the quarkonium mass scale up to the jet scale to resum the leading logarithms of $p_TR/m_c$. The FJF, which obeys a similar evolution equation, is evolved from the jet scale $p_T R$ to the hard scale $p_T$ to resum the logarithms of $R$.

The short-distance coefficients, $d_{g\rightarrow ^3S^{[8]}_1 }(z)$ and $d_{g\rightarrow ^3P^{[8]}_J }(z)$~\cite{Braaten:1994vv,Braaten:2000pc,Braaten:1996rp}, exhibit divergences in the threshold limit as $z\rightarrow 1$. These singularities are tamed by resumming soft gluon emissions through threshold resummation~\cite{Chung:2024jfk}.
At next-to-leading order (NLO) accuracy, the leading-logarithmic (LL) resummed short-distance coefficients are expressed in Mellin space as
\begin{align}\label{eq:FF-LL}
\tilde{d}_{g\to [Q\bar Q(\mathcal{N})]}^{\text{LL}}(N_h)
= 
\exp[J_{\mathcal{N}}(N_h)]
\times 
\left(
\tilde{d}_{g\to [Q\bar Q(\mathcal{N})]}^{\text{LO+NLO}}(N_h)
-
J_{\mathcal{N}}(N_h)\tilde{d}_{g\to [Q\bar Q(\mathcal{N})]}^{\text{LO}}(N_h)
\right)
\end{align}
where $J_{\mathcal{N}}(N_h)$ is given in Ref.~\cite{Chung:2024jfk} and the functions $\tilde{d}$ with a tilde are obtained via Mellin transformation as
\begin{align}
\tilde{d}_{ g\to [Q\bar Q(\mathcal{N})]}(N_h) =
\int^1_0 dz_{J/\psi}\, z_{ J/\psi}^{N_h-1}
d_{ g\to [Q\bar Q(\mathcal{N})] }(z_{J/\psi})\,,
\end{align}
The last term in parentheses in \eq{FF-LL} subtracts the double-log term from the NLO correction to avoid double counting similar to the eventshape resummation \cite{Ee:2025scz,Zhu:2023oka,Chu:2022jgs,Zhu:2021xjn,Chung:2019ota,Kang:2017cjk,Kang:2015swk,Kang:2014qba,Kang:2013nha}. Thus, the threshold resummation for the $^3S_1^{[8]}$ and $^3P_J^{[8]}$ channels is implemented by inserting $\tilde{d}_{g\to [Q\bar{Q}(\mathcal{N})]}^{\text{LL}}$ into \eq{FF-fac}. This process resums the double logarithms of the form $\ln(1 - z_{J/\psi})/(1 - z_{J/\psi})$ and resolves the catastrophic breakdown of gluon FFs in the threshold limit.

Since the FFs from light and heavy quarks into quarkonium differ and provide distinct contributions to the final cross section, the flavor-singlet DGLAP evolution in \eq{DGLAP} alone may not be sufficient. Therefore, nonsinglet evolution for heavy-quark FFs is required to capture these differences.  Following the procedure outlined in~\cite{Vogt:2004ns,Anderle:2015lqa}, the cross section can be decomposed into flavor-singlet (denoted by $\text{S}$) and nonsinglet components (denoted by $q$ for light quark, $c$ for charm quark, and $b$ for bottom quark), such that the singlet and nonsinglet evolutions decouple, yielding 

\bal\label{eq:cs2}
\frac{d\sigma^{ep\rightarrow (\text{jet}J/\psi)X}}{dp_T d\eta dz_{J/\psi}}
= &
\frac{1}{2n_f}\hat{\sigma}_\text{S}
\otimes
\mathcal{G}^{J/\psi}_\text{S} 
 +
\hat{\sigma}^g
\otimes
\mathcal{G}^{J/\psi}_g 
 +
\hat{\sigma}^{q}_{\text{NS}}
\otimes
\mathcal{G}^{J/\psi}_{\text{NS},q} 
 +
\hat{\sigma}^{c}_{\text{NS}}
\otimes
\mathcal{G}^{J/\psi}_{\text{NS},c} 
 +
\hat{\sigma}^{b}_{\text{NS}}
\otimes
\mathcal{G}^{J/\psi}_{\text{NS},b}  \,,
\eal
where $\hat{\sigma}^i$ is a shorthand notation for the PPCS $d\hat{\sigma}_{ep\rightarrow iX}/(dp_T d\eta)$ and the symbol $\otimes$ represents the convolution integral in \eq{cs1}. The singlet partonic cross section is given by the sum over all quark flavors, $\hat{\sigma}^\text{S} = \sum_{f=1}^{n_f} (\hat{\sigma}_{q_f} + \hat{\sigma}_{\bar{q}_f})$, and the singlet FJF $\mathcal{G}^{J/\psi}_\text{S}$ is defined analogously. The nonsinglet partonic cross sections are defined by
\bal
\hat{\sigma}^{c}_{\text{NS}}
= 
-\frac{1}{2n_f}\hat{\sigma}_{\text{S}}
+
\frac{1}{2}
\hat{\sigma}_{c+\bar{c}}
\,,\quad
\hat{\sigma}^{b}_{\text{NS}}
= 
-\frac{1}{2n_f}\hat{\sigma}_{\text{S}}
+
\frac{1}{2}
\hat{\sigma}_{b+\bar{b}}
\,, \quad
\hat{\sigma}^{q}_{\text{NS}}
= 
-\hat{\sigma}^{c}_{\text{NS}}
-\hat{\sigma}^{b}_{\text{NS}}\,.
\eal
where $\hat{\sigma}_{q_f+\bar{q}_f}$ is a short-hand notation for the sum $\hat{\sigma}_{q_f}+\hat{\sigma}_{\bar{q}_f}$. Here, $\hat{\sigma}^{c}_{\text{NS}}$ (and $\hat{\sigma}^{b}_{\text{NS}}$) represent the difference between the charm (and bottom) quark cross section and the averaged partonic cross section, while $\hat{\sigma}^{q}_{\text{NS}}$ corresponds to the difference for the light quarks. This construction ensures that the sum over all flavors vanishes: $\hat{\sigma}^{q}_{\text{NS}} + \hat{\sigma}^{c}_{\text{NS}} + \hat{\sigma}^{b}_{\text{NS}} = 0$. The nonsinglet FJFs $\mathcal{G}^{J/\psi}_{\text{NS},q_f}$ are defined analogously. At LL accuracy, the resummed nonsinglet FJFs can be expressed in Mellin space as
\begin{align}\label{eq:tG}
\tilde{\mathcal{G}}^{J/\psi}_{\text{NS},q}
(N,N_h,p_TR,p_T)
\! = \!
\left(
\frac{\as(p_TR)}{\as(p_T)}
\right)^{{2 \tilde{P}_{qq}(N)}/{\beta_0}}
\!
\left(
\frac{\as(2m_c)}{\as(p_TR)}
\right)^{2 \tilde{P}_{qq}(N_h)/\beta_0}
\!
\tilde{D}^{J/\psi}_{q+\bar{q}}(N_h,2m_c),
\end{align}
where $\beta_0 = \frac{11}{3}N_c - \frac{2}{3}n_f$ is the coefficient of the beta function and $N_c=3$ is the number of colors. The Mellin-space variables $N$ and $N_h$ are conjugate to $z$ and $z_{J/\psi}$, respectively, and $\tilde{P}_{qq}$ is the splitting function in Mellin space. The first exponential factor on the right-hand side represents the evolution of the nonsinglet FJF from the jet scale $p_T R$ to the hard scale $p_T$. The second exponential factor accounts for the evolution of the nonsinglet FF from the mass scale $2m_c$ up to the jet scale $p_T R$. Finally, the singlet FJF $\mathcal{G}^{J/\psi}_S$ and gluon FJF $\mathcal{G}^{J/\psi}_g$ obey the evolution equations in \eq{DGLAP}.

The Mellin-space FJF in \eq{tG} presents a numerical subtlety during its inverse Mellin transformation. This arises from the contributions of the diagonal splitting functions, $P_{qq}$ and $P_{gg}$, during the evolution, whose behavior can render the transformation unstable or even divergent as $z \to 1$. While this subtlety can be treated analytically by reorganizing the $z$-space integration~\cite{Bodwin:2015iua}, we avoid such complications by performing the convolution in \eq{cs2} entirely in Mellin space and only then applying the inverse transformation to obtain the $z_{J/\psi}$ distributions.

Our FJFs at NLO consist of two components: first, the NLO matching coefficients $\mathcal{J}_{ij}$ up to $\mathcal{O}(\alpha_s)$ and second, the NLO FFs, which include $\mathcal{O}(\alpha_s^2)$ terms for all channels, except for the $^3S^{[8]}_1$ channel where both $\mathcal{O}(\alpha_s)$ and $\mathcal{O}(\alpha_s^2)$ terms are incorporated. The PPCSs are evaluated at NLO for both direct and resolved processes. By combining these fixed-order ingredients with both collinear and threshold resummations, our predictions achieve NLO+LL$_{\text{coll}}$+LL$_{\text{thr}}$ accuracy. 

\begin{figure}[tbh!]
	\begin{center}		\includegraphics[width=.95\textwidth]{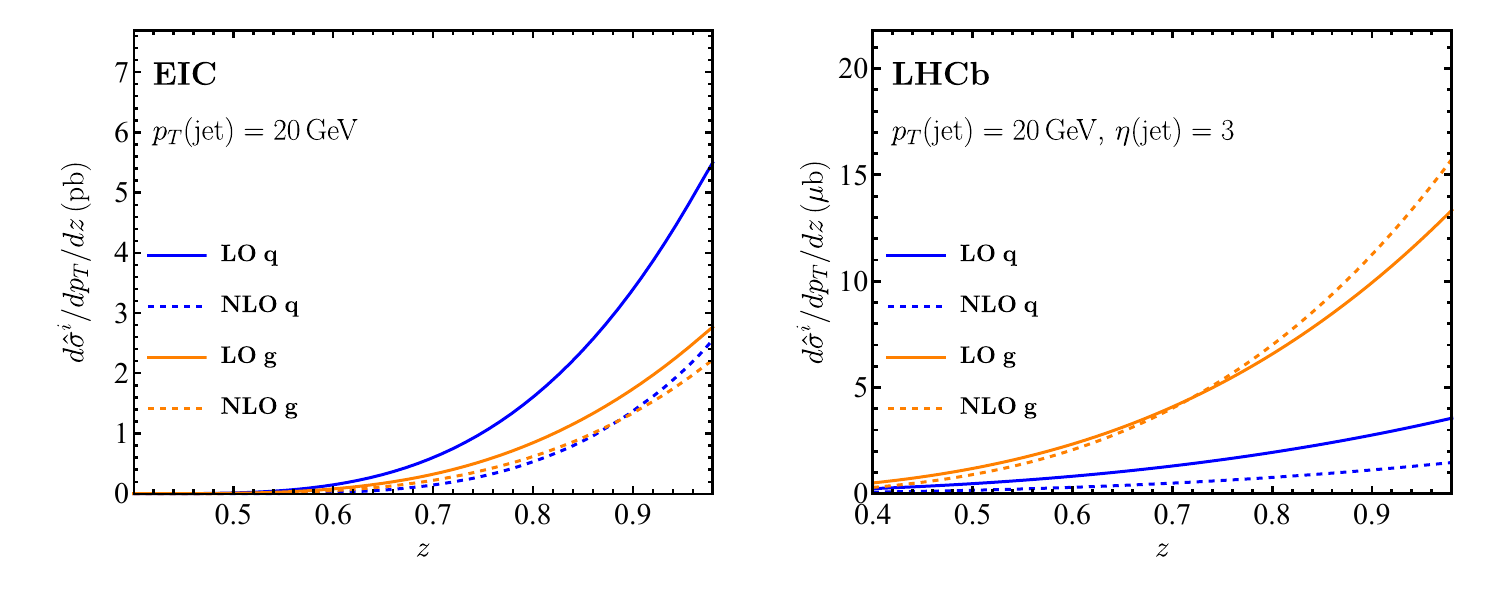}
	\end{center}\vspace{-4ex}
	\caption{LO and NLO PPCSs for quark singlet ($i=q$) and gluon ($i=g$) channels as a function of the momentum fraction $z$, comparing the EIC (left) and LHCb (right).}
	\label{fig:ppcs1}
\end{figure}

\section{Numerical results}
\label{sec:3}
This section presents the numerical results of our analysis. We first detail all  essential input materials and configurations used, including the PPCSs with necessary parameters, the adopted LDME sets, the treatment of feeddown contributions, and the muon cut selections that define the experimental phase space. Subsequently, we present the transverse momentum distributions in comparison to the LHCb results and discuss the features of the EIC results.

\subsection{Setup}
\label{sec:3-1}
We employ the EPHOX Fortran code~\cite{Fontannaz:2001ek, Fontannaz:2001nq, Fontannaz:2002nu, Fontannaz:2003yn} to compute the PPCSs up to NLO for both direct and resolved processes. Our calculation utilizes the CTEQ6 proton PDFs~\cite{Pumplin:2002vw} and the $\text{AFG04}\_\text{BF}$ photon PDFs~\cite{Aurenche:2005da}, with the hard scale set to $p_T$.

The kinematic configuration follows the EIC design specifications~ \cite{AbdulKhalek:2021gbh}. The CM energy of $ep$ collisions is $\sqrt{s}=141\,\text{GeV}$ and the laboratory frame pseudorapidity range is $-2<\eta_{\text{lab}}<4$. The electron and proton beam energies are $E_e=18$ GeV and $E_p=275$ GeV, respectively. To ensure photoproduction, a cut on the virtuality of the photon is imposed, $Q^2<Q^2_{\text{max}}=1\,\text{GeV}^2$. The photon momentum fraction is constrained to $0.2<y<0.8$ to optimize event selection and detector efficiency. This $y$ range corresponds to a $\gamma p$ invariant mass range of $W=\sqrt{(p_\gamma + p_p)^2}$ between 63 GeV and 126 GeV for the EIC kinematics~\cite{Aschenauer:2019uex}.

Fig.~\ref{fig:ppcs1} demonstrates LO and NLO PPCSs for $J/\psi$ production via fragmentation from partons (quark singlet ($i=q$) and gluon ($i=g$)) at the EIC and LHCb. In the EIC case (left panel), the PPCSs exhibit a significant quark contribution (blue lines), while the gluon contribution (orange lines) remains notable. 
In contrast, the LHCb case (right panel) shows a clear dominance of the gluon channel. This is expected because the kinematic region probed by LHCb is highly sensitive to the low $x$ region in $pp$ collisions, where the gluon contribution is dominant.

\begin{figure}[tbh]
	\begin{center}		\includegraphics[width=.95\textwidth]{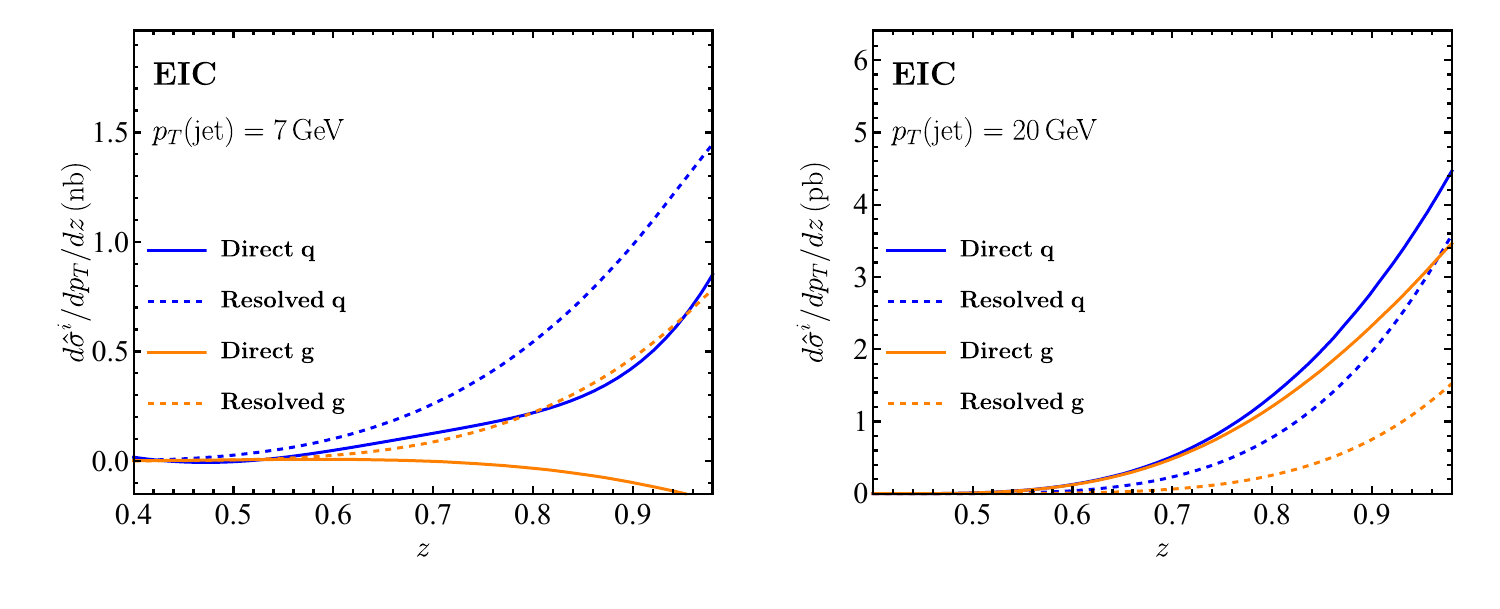}
	\end{center}\vspace{-4ex}
    \caption{The PPCSs for quark singlet ($i=q$) and gluon ($i=g$) channels at the EIC, illustrating the decomposition into direct and resolved components at low $p_T\text{(jet)}=7$ GeV (left) and high $p_T\text{(jet)}=20$ GeV (right) as a function of the momentum fraction $z$.}
	\label{fig:ppcs2}
\end{figure}
Fig.~\ref{fig:ppcs2} provides a decomposition of the EIC PPCSs, separating the contribution of the quasi-real photon $\gamma^*$ interaction into its direct (coupling directly to a quark) and resolved (acting as a source of partons for the hard scattering) components.
In the low-$p_T$ configuration (left panel), the resolved contribution is seen to be larger than the direct component in each parton channel, respectively. This behavior reflects the enhanced sensitivity to the photon's partonic structure at lower transverse momentum, where the effective photon PDFs play a more significant role. In contrast, at higher $p_T$ (right panel), the direct component becomes consistently larger over most of the $z$ range. This trend highlights how the relative importance of the two mechanisms evolves with transverse momentum scale.
Furthermore, the PPCSs for both collision systems exhibit approximately $1/p_T ^5$ or $1/p_T^6$ scaling. This scaling, which causes the PPCS to be dominated by the low-$p_T$ region, in turn reduces the sensitivity at high $p_T$, as indicated by the scale difference in the y-axes between the two panels in Fig.~\ref{fig:ppcs2}.

As an essential step towards the full description of $J/\psi$ production within the FJF framework, partonic results must be convoluted with the FJF, incorporating the LDMEs, $\langle \mathcal{O}^{J/\psi}_{[Q\bar Q(\mathcal{N})]} \rangle$, which describes the evolution of the $Q\bar{Q}(\mathcal{N})$ into $J/\psi$. We consider three representative LDME sets, following the analysis of \cite{Wang:2025drz}. These sets span the three categories identified in global NRQCD analyses, which differ significantly in the relative magnitudes and signs of the color-octet LDMEs.

\begin{table}[thb]
	\centering
	\caption{Representative $J/\psi$ LDME sets used in the calculation, provided in units of GeV$^3$ and $m_c=1.5$~GeV. 
        }
	\label{tab:ME}
	\begin{tabular}{cccccc}
		\hline
		 &\multirow{2}{*}{\phantom{Empty}} 
		& $ \langle \mathcal{O}(^{3}S^{[1]}_{1}) \rangle$ 
		& $ \langle \mathcal{O}(^{3}S^{[8]}_{1}) \rangle$ 
		& $\langle \mathcal{O}(^{1}S^{[8]}_{0}) \rangle$ 
		& $\langle \mathcal{O}(^{3}P^{[8]}_{0}) \rangle / m_c^2$ \\ 
		&
		&  $\times~\text{GeV}^3$ 
		& $\times10^{-2}~\text{GeV}^3$ 
		& $\times~10^{-2}\text{GeV}^3$ 
		& $\times10^{-2}~\text{GeV}^3$ \\ \hline\hline
		
		& Brambilla et al.\cite{Chung:2024jfk,Brambilla:2022ayc}
		& $1.18 \pm 0.35$ & $1.40 \pm 0.42$ & $-0.63 \pm 3.22$ & $2.33 \pm 0.83$ \\ \cline{2-6} 
		
		& Bodwin et al. \cite{Bodwin:2015iua}
		& $1.32 \pm 0.20$ & $-0.71\pm 0.36$ & $11.0 \pm 1.4$ & $-0.31 \pm 0.15$ \\ \cline{2-6}
		
		& B\&K  \cite{Butenschoen:2011yh} 
		& $1.32 \pm 0.20$ & $0.22 \pm 0.06$ & $4.97 \pm 0.44$ &$-0.72 \pm 0.09$ \\ \hline\hline
	\end{tabular}
\end{table}

The three general classes of LDME sets available in the literature are
\begin{itemize}
\item{
\textit{Category 1}: The ${}^3S_1^{[8]}$ and ${}^3P_J^{[8]}$ terms have the same sign and providing the dominant contribution to the inclusive cross section, while the ${}^1S_0^{[8]}$ is small.
Refs.~\cite{Chung:2024jfk,Brambilla:2021abf,Han:2014jya,Zhang:2014ybe} and Ref.~\cite{Shao:2014yta} (minimized ${}^1S_0^{[8]}$) fall into this category.
}

\item{
\textit{Category 2}: The ${}^3S_1^{[8]}$ and ${}^3P_J^{[8]}$ terms retain the same sign. In this category, ${}^1S_0^{[8]}$ is large and dominates the inclusive cross section.  Refs.~\cite{Gong:2012ug,Bodwin:2015iua,Feng:2018ukp}, Table II (fit D) of Ref.~\cite{Butenschoen:2022qka}, and Ref.~\cite{Shao:2014yta} (with maximized ${}^1S_0^{[8]}$) belong here.
}

\item{
\textit{Category 3}: The ${}^3S_1^{[8]}$ and ${}^3P_J^{[8]}$ terms have opposite signs (positive and negative, respectively).  Ref.~\cite{Butenschoen:2011yh} and fits A and B of Ref.~\cite{Butenschoen:2022qka} fall into this category.
}
\end{itemize}
Representative sets of LDME are listed in \tab{ME}, where the predictions across sets within the same category are expected to be similar.

The prompt $J/\psi$ production consists of the direct process\footnote{Here, `direct' refers to $J/\psi$ production from partonic processes excluding feeddown from excited states. This should be distinguished from the 'direct' process in photoproduction.}, where a parton fragments into $J/\psi$, and feeddown process from excited charmonium states $\psi(2S)$, $\chi_{c1}$ and $\chi_{c2}$, assuming that non-prompt $J/\psi$ contributions from $B$ decay are well excluded. We adopt the corresponding LDMEs for these states from Refs.~\cite{Chung:2024jfk,Brambilla:2022ayc,Bodwin:2015iua}, and use their respective branching ratios, $\{0.615,\; 0.343,\; 0.195\}$, to convert these states to the final $J/\psi$ yield. For phase space corrections arising from quarkonium mass differences, we implement kinematic rescaling through the compensation factor $(m_H/m_{J/\psi})z_{J/\psi}$, consistent with the procedure in Ref.~\cite{Bodwin:2015iua}. The treatment of the B\&K set differs from the others, as the feeddown contributions are implicitly absorbed in its fitted LDME values according to the original prescription.

Comparing theoretical predictions to experimental data requires accounting for the measurement's kinematic setup.
Since $J/\psi$ is usually reconstructed via its muon-pair decay channels, the experimental acceptance is primarily determined by the stringent muon identification requirements of each detector. The muon identification requirements for the relevant experiments are as follows:
\begin{itemize}
\item{
LHCb~\cite{LHCb:2017llq}: Both muons satisfy $2.0 < \eta(\mu) < 4.5$ and $p_T(\mu) > 0.5$ GeV, with the additional requirement that $p(\mu) > 5$ GeV, and $\sqrt{p_T(\mu^+) p_T(\mu^-)}>\text{1.5 GeV}$.}

\item{
CMS~\cite{CMS:2021znk, CMS:2021mzx}: Each muon satisfies one of the following kinematic domains, depending on its pseudorapidity: $p_T(\mu)>3.5~\text{GeV}$ for $|\eta(\mu)|<1.2$; $p_T(\mu)>(5.47-1.89|\eta(\mu)|)~\text{GeV}$ for $1.2<|\eta(\mu)|<2.1$; and $p_T(\mu)>1.5~\text{GeV}$ for $2.1<|\eta(\mu)|<2.4$.
}
\end{itemize}

\begin{figure}[tbh]
    \begin{center}	
    \includegraphics[width=.5\textwidth]{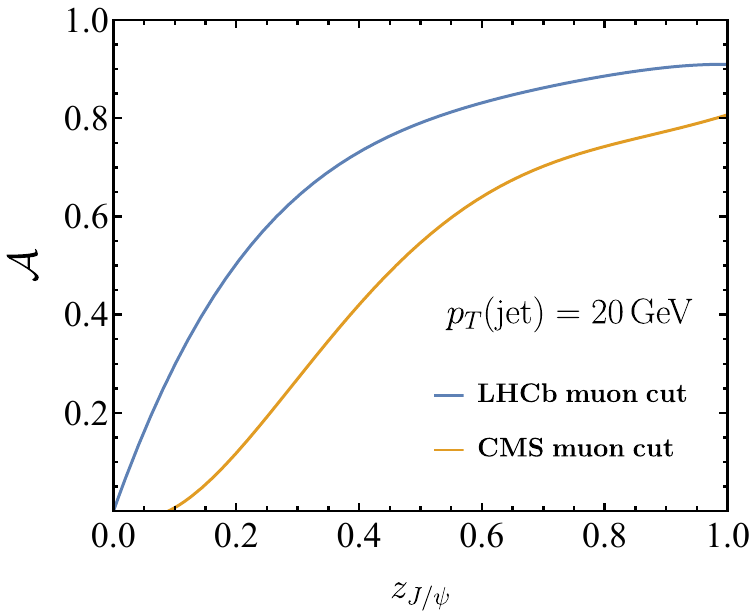}
	\end{center}\vspace{-4ex}
	\caption{The muon acceptance, $\mathcal{A}$, as a function of the momentum fraction $z_{J/\psi}$ calculated for LHCb and CMS kinematic regions.}
	\label{fig:muon_cut}
\end{figure}

The muon acceptance, $\mathcal{A}$, calculated under the specific kinematic cuts for both the LHCb and CMS experiments is illustrated in Fig.~\ref{fig:muon_cut}. The LHCb acceptance (blue) yields higher acceptance, exceeding $80\%$ for $z_{J/\psi} \approx 0.5$, while the CMS acceptance (orange) 
is generally lower across the entire $z_{J/\psi}$ range. This difference stems directly from the variations in each experiment's pseudorapidity acceptance and the applied momentum threshold, resulting in a severe deformation of the predicted $z_{J/\psi}$ distribution. This suppression is particularly pronounced at small $z_{J/\psi}$. Therefore, this kinematic bias must be carefully corrected to obtain accurate physics results.

For our study at the EIC, where the final muon identification criteria are not yet established, we adopt the LHCb muon cuts as a baseline, given its typically higher acceptance and broader $z_{J/\psi}$ coverage. We adjust the pseudorapidity range to $-2 < \eta < 4$ to better match the expected angular coverage of the EIC detector. These results should be regarded as preliminary and will be updated once the EIC muon selection criteria are established.

\subsection{Transverse momentum fraction distributions}
\label{sec:3-2}
In this subsection, we present the $z_{J/\psi}$ distributions obtained using the three LDME sets listed in \tab{ME}. We discuss the key features of the EIC predictions in comparison to LHCb results and examine how these distributions depend on the jet radius $R$.

\begin{figure}[tbh]
	\begin{center}
		\includegraphics[width=.3\textwidth]{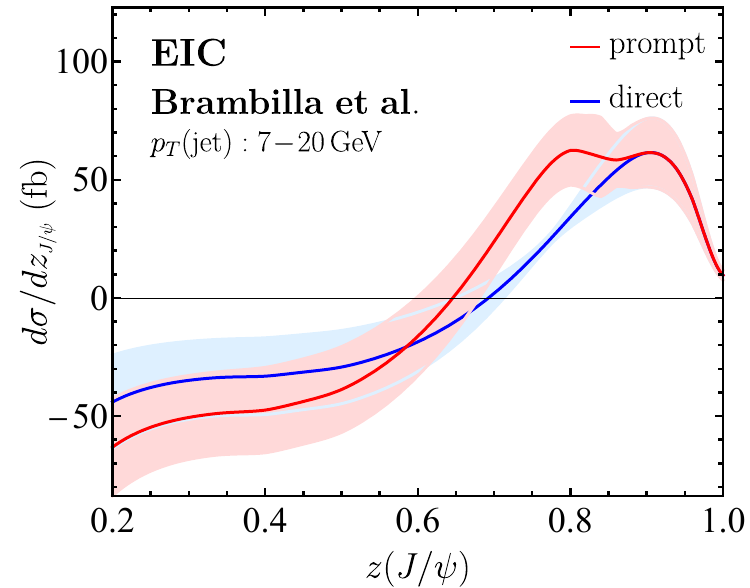}
		\includegraphics[width=.3\textwidth]{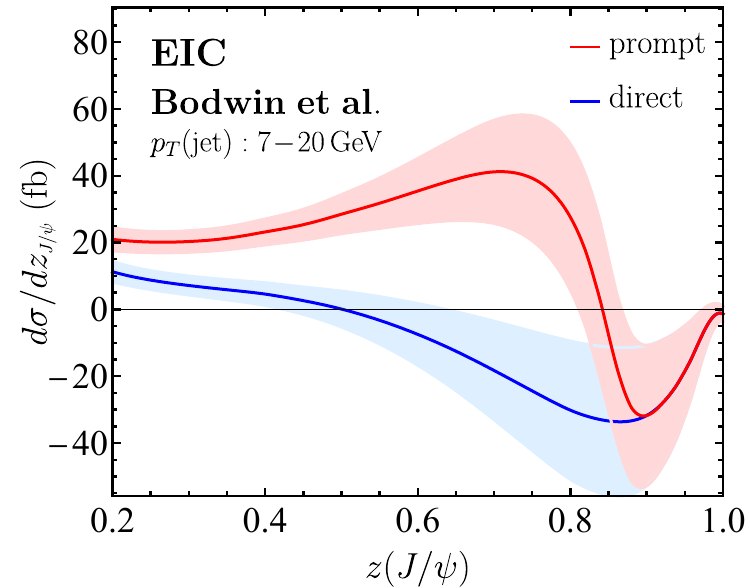}
		\includegraphics[width=.3\textwidth]{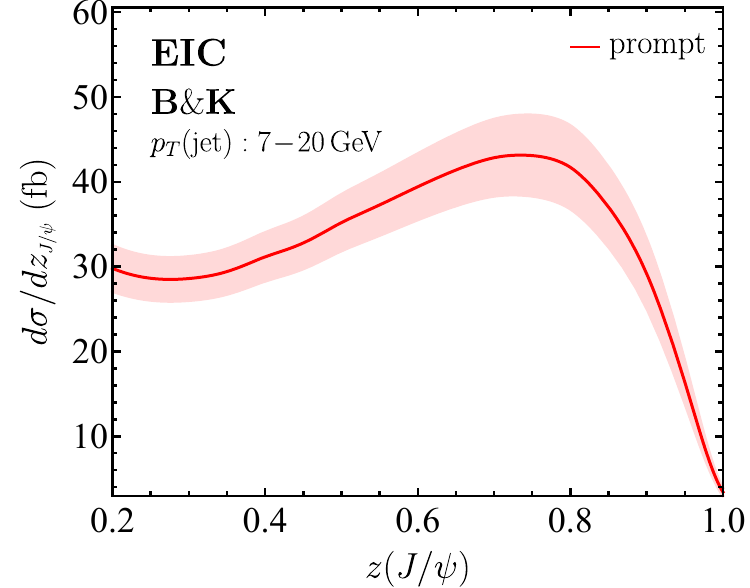}
		\includegraphics[width=.3\textwidth]{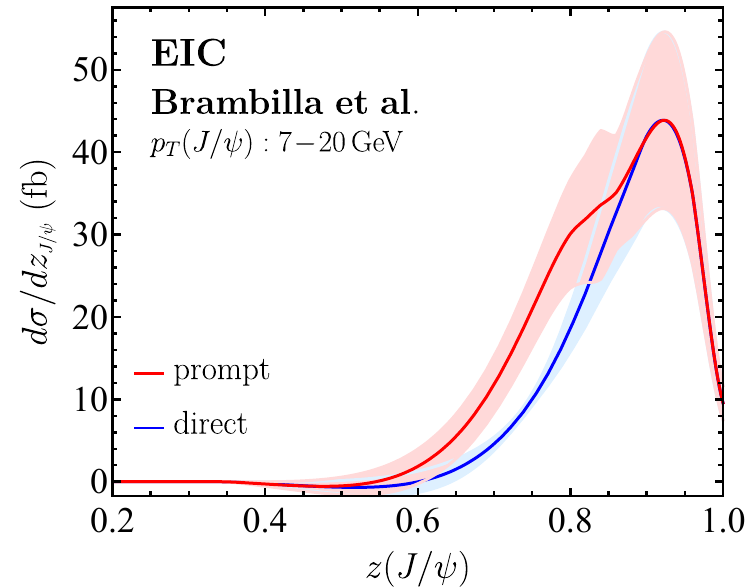}
		\includegraphics[width=.3\textwidth]{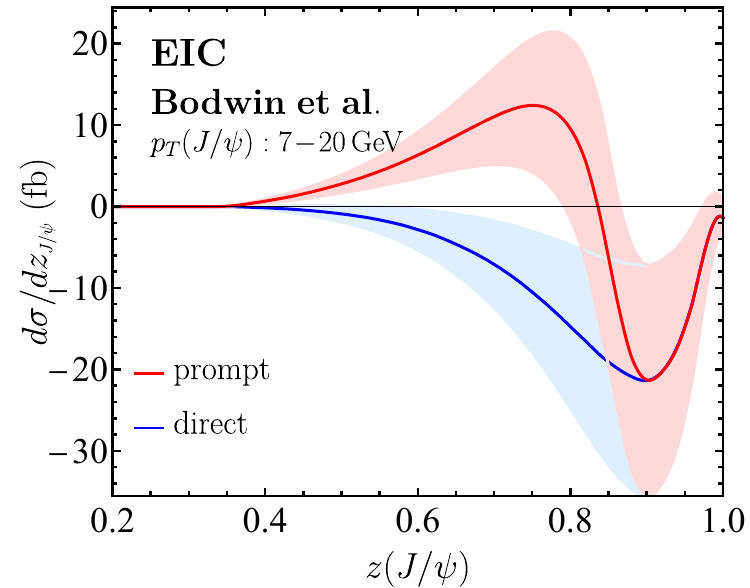}
		\includegraphics[width=.3\textwidth]{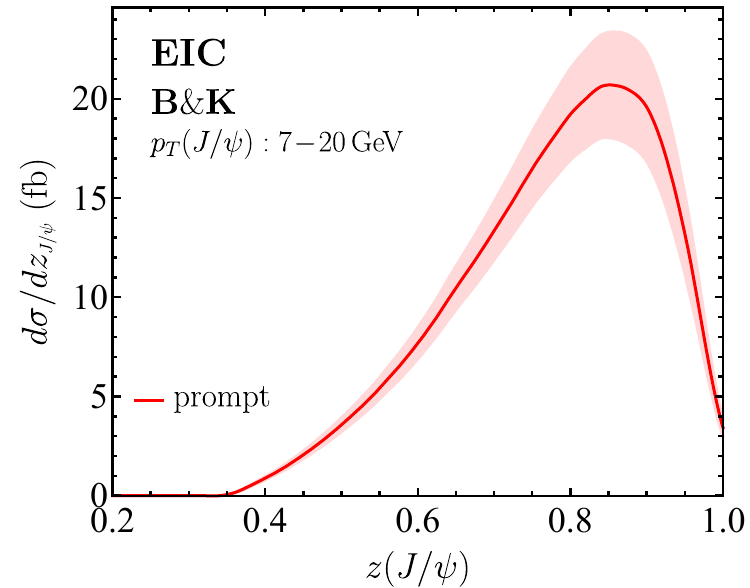}
	\end{center}\vspace{-4ex}
	\caption{Predicted $z_{J/\psi}$ distributions for three sets of LDMEs in \tab{ME}, with the upper and lower panels displaying results in jet-$p_T$ bin and $J/\psi$-$p_T$ bin respectively.}
	\label{fig:jpsi_EIC_1}
\end{figure}

Fig.~\ref{fig:jpsi_EIC_1} shows $z_{J/\psi}$ distributions in jet-$p_T$ (upper) and $J/\psi$-$p_T$ (lower) bins, each within the range $7 < p_T < 20$ GeV.{For clarity, we denote the jet transverse momentum as $p^{\text{jet}}_T$ from now on.}
Jets are reconstructed via the anti-$k_T$ algorithm \cite{Cacciari:2008gp} using a radius parameter $R=0.8$~\cite{Aschenauer:2019uex}. To facilitate a clearer comparison of the LDME effects, we calculate the theoretical uncertainties using the full covariance matrices provided for the three LDME sets.

The prompt distributions (red curves) represent the sum of direct and feeddown contributions. In the upper panel, all three sets produce a peak in the large-$z_{J/\psi}$ region but yield distinct shapes across the entire range, highlighting the discriminating power of the FJF framework. The Brambilla set exhibits an unphysical negative distribution in the small-$z_{J/\psi}$ region because the large negative ${}^3P_J^{[8]}$ contribution overwhelms the positive ${}^3S_1^{[8]}$ contribution. Conversely, the Bodwin set yields a negative distribution in the large-$z_{J/\psi}$ region, as the large negative ${}^3S_1^{[8]}$ contribution outweighs the positive ${}^3P_J^{[8]}$ component. For the B\&K set, only the prompt distribution is presented—since feeddown contributions are presumed to be absorbed into the LDME values—and it remains positive over the entire $z_{J/\psi}$ range.
The direct distribution (blue curve) is negative across most of the $z_{J/\psi}$ range for the Bodwin set, which indicates a dominant feeddown contribution. For the Brambilla set, the feeddown contribution accounts for roughly 35\% of the total yield, implying that the direct and feeddown contributions are comparable in magnitude.

The lower boundary of 7 GeV (which is more than twice the $J/\psi$ mass) is chosen to suppress power corrections of order $(m/p_T^{J/\psi})^2$ that are not accounted for in the factorization formula in \eq{cs1}, which was derived in the massless limit.
The lower panel also implements an upper bound of $p_T^{\text{jet}} < 20$ GeV, which causes the distributions to vanish for $z_{J/\psi} < 0.35$. 

Regarding the muon cut effects discussed in Sec.~\ref{sec:3-1}, the muon acceptance significantly distorts the distribution at small $z_{J/\psi}$ in the upper panel; specifically, the application of the CMS muon selection would result in a substantial distortion. In contrast, the lower panel is less sensitive to the muon cuts because the small-$z_{J/\psi}$ region is already kinematically suppressed by the upper limit on $p_T^{\text{jet}}$, according to the definition $z_{J/\psi} = p_T^{J/\psi}/p_T^{\text{jet}}$.

\begin{figure}[tbh]
	\begin{center}
		\includegraphics[width=.4\textwidth]{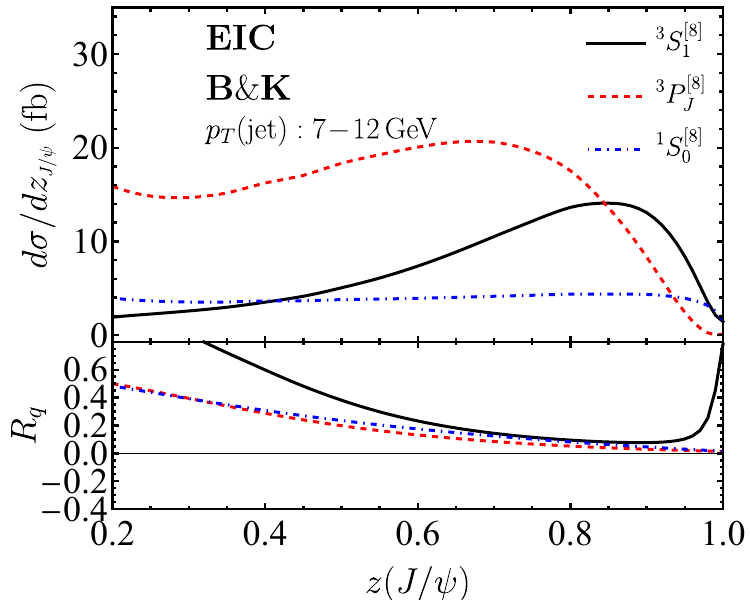}
		\includegraphics[width=.4\textwidth]{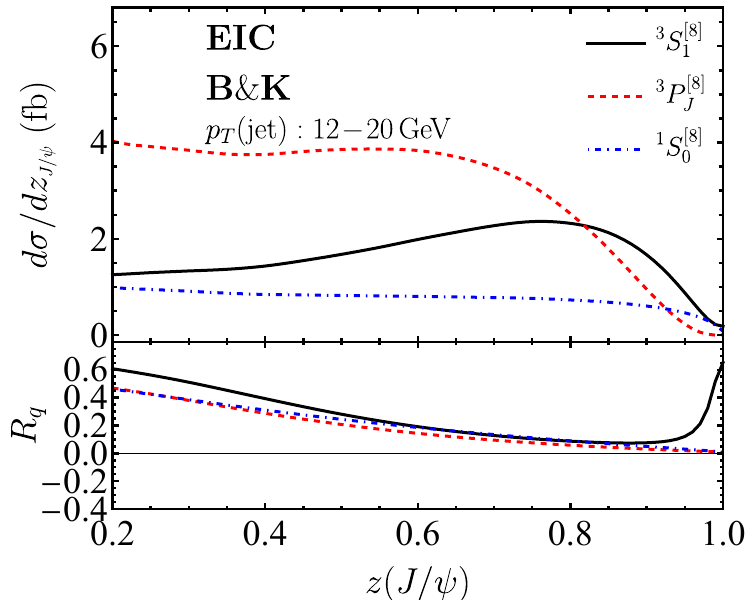}
		\includegraphics[width=.4\textwidth]{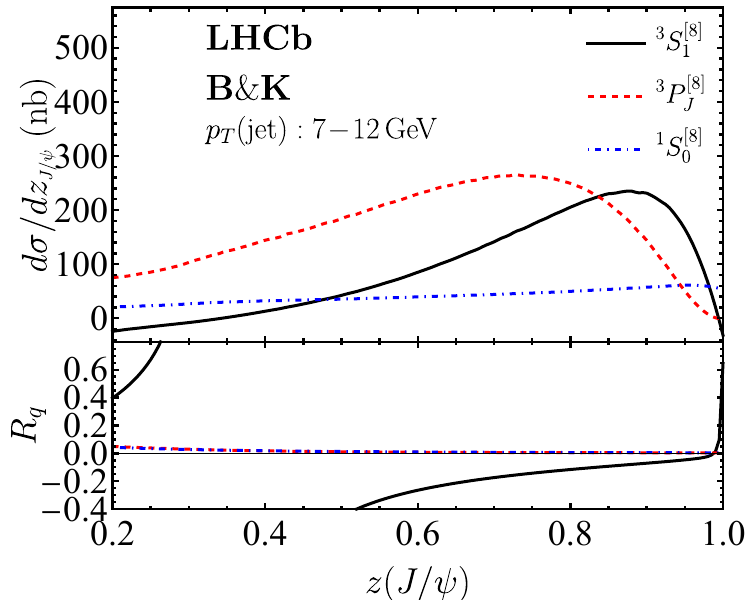}
		\includegraphics[width=.4\textwidth]{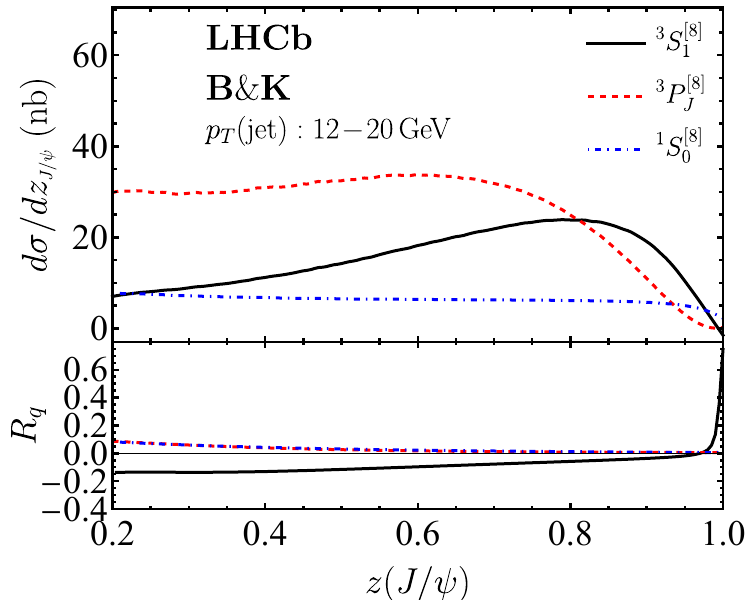}
	\end{center}\vspace{-4ex}
	\caption{Predicted $z_{J/\psi}$ distributions and octet-state fractions for the EIC (upper panel) and LHCb (lower panel). }
	\label{fig:jpsi_EIC_2}
\end{figure}

The distributions in the upper panel of \fig{jpsi_EIC_1} closely resemble the corresponding LHCb results from Ref.~\cite{Wang:2025drz}, except in the small-$z_{J/\psi}$ region.
The enhanced quark PPCS at the EIC in \fig{ppcs1} implies distinct distributions from those of the LHCb. However, the quark-initiated FFs largely compensate for this enhancement.
In $^3P^{[8]}_J$ and $^1S^{[8]}_0$ channels, while their gluon FFs arise at NLO ($\alpha_s^2$), quark FFs starting at NNLO ($\alpha_s^3$) are currently unknown, thus quark contributions to these channels are negligible.
For the $^3P^{[8]}_J$ and $^1S^{[8]}_0$ channels, while their gluon FFs arise at NLO ($\alpha_s^2$), the quark FFs—which start at NNLO ($\alpha_s^3$)—are currently unknown. Thus, the quark contributions to these channels are negligible.
The $^3S^{[8]}_1$ channel possesses a gluon FF starting at LO ($\alpha_s$). At NLO ($\alpha_s^2$), the gluon FF remains significantly larger than the quark FF, ultimately leading to suppressed quark contributions to the total yield.
The $^3S^{[1]}_1$ channel, which receives a contribution from charm-quark fragmentation starting at NLO ($\alpha_s^2$), is suppressed due to small heavy-quark PPCSs.

To explore the quark contribution in the small-$z_{J/\psi}$ region, we plot the individual contributions from the color-octet channels for two jet-$p_T$ bins in Fig.~\ref{fig:jpsi_EIC_2} for both the EIC (upper panel) and the LHCb (lower panel), taking the B\&K set as a representative example. For the B\&K set, the $^3\!S^{[8]}_1 + ^3\!P^{[8]}_J$ channels dominate the cross sections in both cases. To quantify the quark contribution, we define the quark fraction $R_q$ as:
\begin{align}\label{eq:Rq}
R_q
= 
\frac{ 
\sum_{j\in\{q\}}
\hat{\sigma}_{ep\to jX} \otimes
\mathcal{G}_{j}^{J/\psi}}{\mathop{\sum}_{j\in\{q,g\}}
\hat{\sigma}_{ep\to jX} \otimes
\mathcal{G}_j^{J/\psi}},
\end{align}
and display the corresponding distribution in the lower part of each plot.
The quark fraction $R_q$ for the EIC is relatively larger than that for the LHCb. $R_q$ increases in the small-$z_{J/\psi}$ region because the contributions from off-diagonal evolution are enhanced via the splitting function $P_{qg}(z_{J/\psi} \to 0)$. The $^3S_1^{[8]}$ contribution, which involves the quark FFs, is more pronounced in the low jet-$p_T$ bin, whereas the $^1S^{[8]}_0$ and $^3P^{[8]}_J$ contributions remain largely unchanged. For the $^3S^{[8]}_1$ channel, $R_q$ grows rapidly as $z_{J/\psi} \rightarrow 1$ due to the $\ln(1-z_{J/\psi})$ enhancement in its quark FF~\cite{Ma:1995vi}. Although threshold resummation for this channel is not currently available, its effect on the $d\sigma/dz_{J/\psi}$ distributions is not expected to be significant. Finally, the LHCb 7–12 GeV jet-$p_T$ bin displays a zero-crossing at $z_{J/\psi} \approx 0.3$, which results in a hyperbolic feature in the corresponding $R_q$ plot.

\begin{figure}[tbh]
	\begin{center}		\includegraphics[width=.4\textwidth]{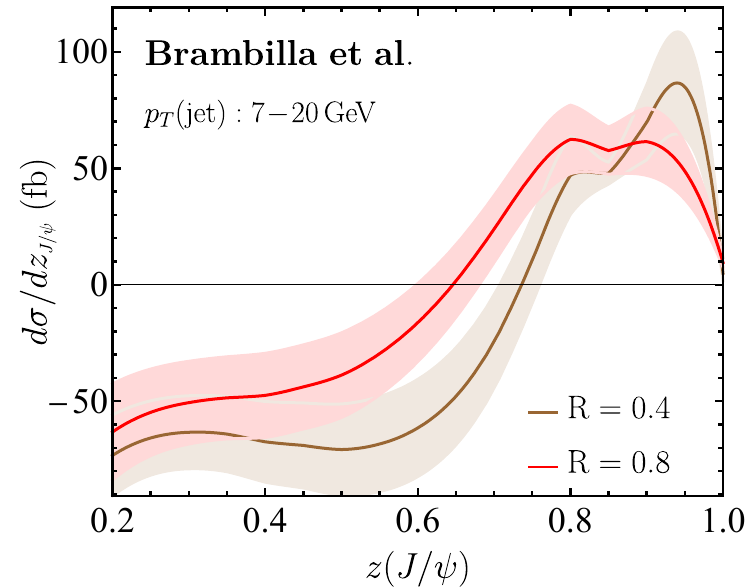}
    \includegraphics[width=.4\textwidth]{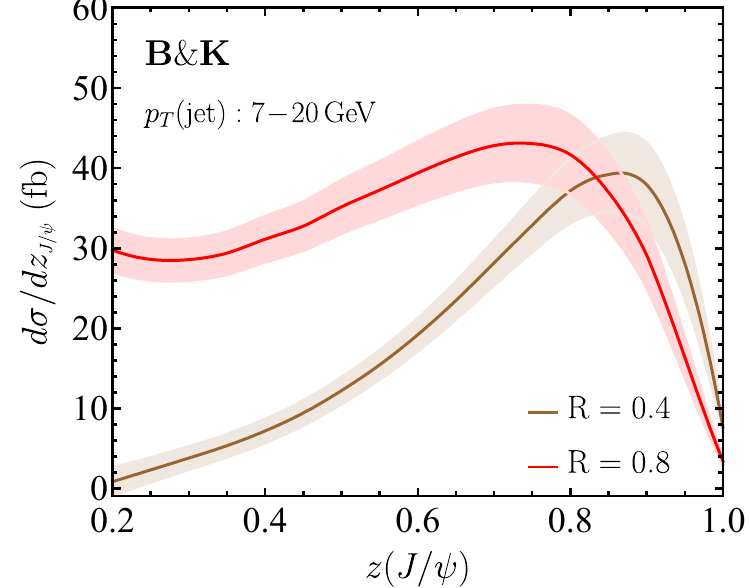}
    \includegraphics[width=.4\textwidth]{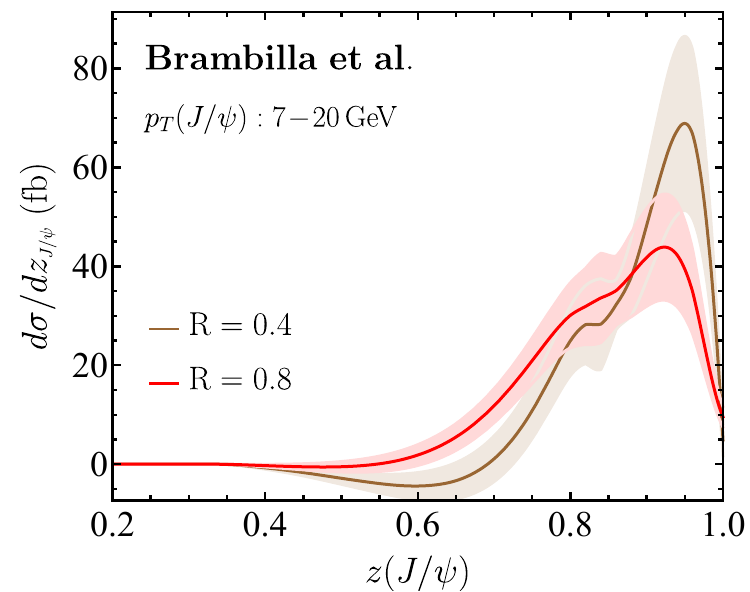}
    \includegraphics[width=.4\textwidth]{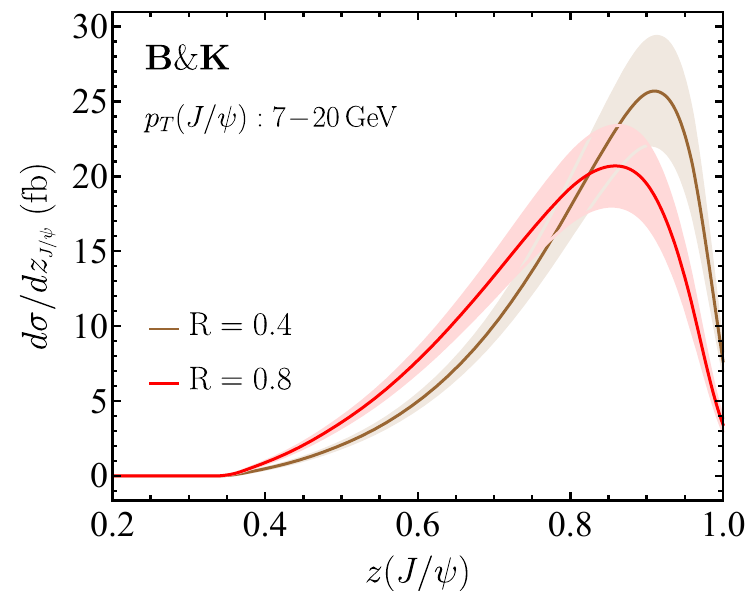}
	\end{center}\vspace{-4ex}
	\caption{Predicted $z_{J/\psi}$ distributions for jet radius $R=0.4$ and $R=0.8$ in jet-$p_T$ bin (upper panel) and $J/\psi$-$p_T$ bin (lower panel).}
	\label{fig:jpsi_EIC_6}
\end{figure}

Fig.~\ref{fig:jpsi_EIC_6} shows the distributions for jet radii $R=0.4$ (brown) and $R=0.8$ (red) in the $7 < p_T^{\text{jet}} < 20$ GeV bin. The radius $R$ enters via the jet coefficients $\mathcal{J}_{ij}$ and the jet scale $p_T R$ associated with the DGLAP evolution. A smaller jet radius, such as $R=0.4$, results in smaller power corrections of order $\mathcal{O}(R^2)$ under the narrow-jet approximation.
The distributions exhibit significant sensitivity to the value of $R$, with a smaller radius yielding a distinctly narrower peak and distorting the small-$z_{J/\psi}$ region of the profile.

\section{Summary}
\label{sec:4}

We examined $J/\psi$ distributions within jets at the EIC to test their potential for discriminating between different production mechanisms and providing complementary information to the results from the LHC.

We gave predictions for $J/\psi$ production within jets at the EIC using the fragmenting jet function framework. The results are presented as differential cross sections $d\sigma/dz_{J/\psi}$, where $z_{J/\psi} = p^{J/\psi}_T / p^{\text{jet}}_T$ is the quarkonium transverse momentum fraction. These cross sections are computed using a factorization formula composed of partonic cross sections and FJFs.
The FJFs encode jet coefficients and heavy quarkonium FFs, which contain the NRQCD long-distance matrix elements. DGLAP evolution is applied to both the FJFs and FFs to resum collinear logarithms, while threshold resummation is incorporated into the FFs to ensure well-behaved and convergent distributions near the kinematic limit $z_{J/\psi} \to 1$.
We implemented the NLO partonic cross sections for direct and resolved photoproduction via the EPHOX package, yielding final distributions with NLO accuracy plus LL resummation of collinear and threshold logs.

We explored the impact of muon selection criteria by applying both LHCb and CMS-like cuts. We found that both sets of criteria significantly distort the distributions, with the degree of distortion being highly sensitive to the specific selection requirements. 
Muon acceptance significantly suppresses the distribution at small $z_{J/\psi}$ in jet-$p_T$ bins, while the effect is reduced in bins of $J/\psi$ $p_T$.
For definiteness, we adopted the LHCb criteria as the baseline for our EIC distribution predictions.
We also investigated the sensitivity to the jet radius $R$ and found that a smaller $R$ yields a narrower peak in the distribution profile.

The EIC will provide independent information and complementary constraints on quarkonium production. Specifically, future measurements of $z_{J/\psi}$ distributions are expected to impose important constraints on various NRQCD long-distance matrix-element sets. Ultimately, a systematic comparison with existing LHC data will further advance our understanding of quarkonium production mechanisms.

\acknowledgments 
DK would like to thank Soohwan Lee for stimulating discussions.
TH, DK, and YW  are supported by the National Natural Science Foundation of China (NSFC) through National Key Research and Development Program under the contract No. 2024YFA1610503. The work of U-R.K. is supported by the National Research Foundation of Korea(NRF) grant funded by the korea government (MSIT) under contract No.~RS-2025-24533579.

\appendix

\clearpage
\bibliographystyle{JHEP}
\bibliography{main.bbl}


\end{document}